\documentclass[letterpaper, 10 pt, conference]{ieeeconf}  

\IEEEoverridecommandlockouts                              

\usepackage{amsmath}
\usepackage{amssymb}
\usepackage{graphicx}
\usepackage{breqn}
\usepackage{url}

\newtheorem{definition}{Definition}
\newtheorem{theorem}{Theorem}
\newtheorem{example}{Example}
\newtheorem{lemma}{Lemma}
\newtheorem{proposition}{Proposition}
\newtheorem{remark}{Remark}

\newcommand{\eq}[1]{\begin{equation*}#1\end{equation*}}
\newcommand{\eql}[1]{\begin{equation}#1\end{equation}}

\newcommand{\EE}{\mathbb{E}}
\newcommand{\PP}{\mathbb{P}}

\newcommand{\ins}[1]{\mathrm{#1}}
\renewcommand{\d}{\ins{d}}
\newcommand{\norm}[1]{\|#1\|}
\newcommand{\bx}{x}
\newcommand{\bff}{f}
\newcommand{\by}{y}

\title{\LARGE \bf
Guaranteed
Control of Sampled Switched Systems 
using Semi-Lagrangian Schemes
and One-Sided Lipschitz Constants 
}

\author{Adrien Le Co\"ent$^{1}$ and Laurent Fribourg$^{2}$
\thanks{$^{1}$Department of Computer Science, Aalborg University
        {\tt\small adrien@cs.aau.dk}}%
\thanks{$^{2}$LSV - ENS Paris-Saclay \& CNRS
        {\tt\small fribourg@lsv.fr}}%
}

\begin{document}

\maketitle
\thispagestyle{empty}
\pagestyle{empty}

\begin{abstract}
In this paper, we propose a new method
for ensuring formally that a controlled
trajectory stay inside a given safety set ${\cal S}$
for a given duration $T$.
Using a finite gridding~${\cal X}$ of ${\cal S}$,
we first synthesize, for a subset of initial nodes~$x$
of ${\cal X}$,
an {\em admissible control}
for which the Euler-based {\em approximate trajectories}
lie in ${\cal S}$ at~$t\in [0,T]$.
We then give sufficient conditions which ensure that
the {\em exact trajectories}, under the same control, 
also lie in ${\cal S}$ for~$t\in [0,T]$, when
starting
at initial points ``close'' to nodes~$x$.
The statement of such conditions relies
on results giving estimates of the deviation
of Euler-based approximate trajectories, using {\em one-sided Lipschitz constants}.
We illustrate the interest of the method on
several examples, including a stochastic one.
\end{abstract}

\section{Introduction}

Consider an ordinary differential equation (ODE) of the form $\dot{z}=f(z)$
on $\mathbb{R}^n$.
Classically, one knows that, if the 
function $f$ 
is Lipschitz continuous with Lipschitz constant~$L$, the solution of the ODE
starting at a given initial value exists and is unique. Besides, one has:
\begin{equation}\label{eq:1}
\|X_{t,z_1}-X_{t,z_2}\| \leq e^{L t}\|z_1-z_2\|,
\end{equation}
where $\|\cdot\|$ denotes the Euclidean norm, and $X_{t,z_i}$ denotes the value of the solution of the ODE at time~$t$, starting at initial value $z_i$ ($i=1,2$).
This gives a rough {\em growth bound}, i.e. a function bounding the distance of neighboring trajectories as $t$ evolves.

In the 90's, several researchers \cite{donchev,lempio} have 
obtained a more accurate growth bound, using the notion of ``one-sided Lipschitz (OSL)'' function. The function~$f$ is said to be OSL if there exists a constant $\lambda\in\mathbb{R}$ such that, for all $z_1,z_2\in\mathbb{R}^n$:
$$\langle f(z_1)-f(z_2), z_1-z_2\rangle \leq \lambda \| z_1 -z_2 \|^2,$$
where $\langle \cdot, \cdot\rangle$ denotes the scalar product of two vectors of $\mathbb{R}^n$.
The real $\lambda$ is called the OSL constant associated with $f$.
In \cite{donchev}, it is proven that, if $f$ is continuous and OSL
with OSL constant~$\lambda$, then the solution of the ODE starting at a given
initial value exists and is unique, and, for all
$z_1,z_2\in\mathbb{R}^n$:
\begin{equation}\label{eq:2}
\| X_{t,z_1}-X_{t,z_2}\| \leq e^{\lambda t}\|z_1-z_2\|.
\end{equation}
This gives a more accurate growth bound because
a Lipschitz function $f$ is always OSL, and the associated OSL constant~$\lambda$ is always less than or equal to its Lipschitz counterpart~$L$.
Furthermore,
in the case of ``stiff'' differential equations, we have $\lambda \ll L$
(see \cite{donchev}). Note also that a function can be OSL but not Lipschitz (not even {\em locally} Lipschitz): inequation (\ref{eq:2}) then still applies
while inequation (\ref{eq:1}) does not apply any longer.
Using the OSL constant $\lambda$, it is also possible to bound the error 
$\| X_{t,z_1}-\tilde{X}_{t,z_2}\|$ in function of $\| z_1-z_2\|$, where $\tilde{X}_{t,z_2}$ denotes the {\em Euler 
approximate} $z_2+tf(z_2)$ of the solution~$X_{t,z_2}$.
In \cite{c13}, we have derived some analytic forms of such error estimates
when one focuses on a compact subdomain
${\cal S}\subset \mathbb{R}^n$ of solutions. We have also 
given an OSL-based error estimate  for (a variant of) the Euler-Maruyama
approximate solution in the case of {\em stochastic} ODEs \cite{tamed}.
These results have been used to synthesize 
controls that are ``correct-by-construction'', in the sense that they are guaranteed to satisfy given {\em safety constraints} \cite{c13,tamed}.
In this paper, we show how such error estimates can be integrated to
{\em semi-Lagrangian (SL) schemes} in order to synthesize 
{\em optimal} controls for problems with safety constraints.


\ 

The plan of the paper is as follows:
in Section II, we present the context of our work
and the principle of the method; in Section III,
we give formal sufficient conditions that guarantee the
safety of the control;
Section IV illustrates on two examples how the method can be extended for stochastic ODEs and differential games;
we conclude in Section V.

\section{Context and Principle of the Method}
Let us present the context and the principle of our method.
\subsection{Switched systems}\label{ss:defs}
A hybrid system is a system where the state evolves continuously according
to several possible modes, and where the change of modes (switching) is done
instantaneously.We consider here the special case of hybrid systems called 
``sampled
switched systems'' \cite{c2} where the change of modes occurs periodically with a
period of $\tau$ seconds. We will suppose furthermore that the state keeps its value
when the mode is changed (no jump). More formally, we denote the state of the
system at time $t$ by $z(t)\in\mathbb{R}^n$. The set of modes $A$ is {\em finite}. With
each mode $a\in A$ is associated a vector field $f_a$ that governs the state 
$z(t)$, we
have:
$$\dot{z}(t)=f_a(z(t)).$$
We make the following hypothesis:

$(H0)$ For all $a\in A$, $f_a$ is a locally Lipschitz continuous map.\\

\ 

We will denote by $X^a_{t,z_0}$ the solution at time $t$ of 
$\dot{z}(t)=f_a(z(t))$ with $z(0)=z_0$.
The existence of $X^a_{t,z_0}$ is guaranteed by assumption $(H0)$. Let us consider ${\cal S}\subset\mathbb{R}^n$ be a compact and convex set, typically a ``rectangular set'', i.e. a cartesian product on $n$ closed intervals. We know by $(H0)$ that there exists a constant $L_a>0$ such that:
\begin{equation}
\|f_a(z_1)-f_a(z_2)\| \leq L_a \|z_1-z_2\|\ \ \ \forall  z_1,z_2\in {\cal S}.
\end{equation}
We also define, for all $a\in A$:
\begin{equation}
C_a = \sup_{z\in {\cal S}} L_a\|f_a(z)\|.
\end{equation}
Let us denote by ${\cal T}$ a compact overapproximation of the set of
trajectories
starting in ${\cal S}$ for $0\leq t\leq \tau$, i.e. ${\cal T}$ is such that
\begin{equation}
{\cal T} \supseteq \{ X^a_{t,z_0}\ |\ a\in A, 0\leq t\leq\tau, z_0\in {\cal S}\}.
\end{equation}
The existence of ${\cal T}$ is guaranteed by assumption $(H0)$. 
It follows from $(H0)$ that the vector fields $f_a$ of the system are OSL on ${\cal T}$: for all $a\in A$, there exists 
a constant $\lambda_a\in\mathbb{R}$ such that
\begin{equation}
\langle f_a(z_1)-f_a(z_2), z_1-z_2\rangle \leq \lambda_a \|z_1-z_2\|^2\ \ \ \forall z_1,z_2\in {\cal T}.
\end{equation}

We consider a {\em finite time horizon} problem: we suppose that time $t$ belongs to interval $[0,k\tau]$, where $k$ is a given integer number.
%
%
Given a sequence of modes (or ``pattern'') $\pi := a_1\cdots a_k\in A^k$, we denote by
$X_{t,z_0}^{\pi}$
the solution of the ODE of mode $a_1$ for $t\in [0,\tau[$
with initial condition~$z_0$,
extended continuously with the solution of the ODE of mode $a_2$ for $t\in[\tau,2\tau[$, and so on iteratively until mode $a_k$ for $t\in[(k-1)\tau,k\tau]$.\\
\subsection{Optimal problems}
We consider the {\em cost function}: $J_{k,\tau}:\mathbb{R}^n\times A^k\rightarrow \mathbb{R}_{\geq 0}$ 
defined by:
$$J_{k,\tau}(z_0,\pi)=\|X_{k\tau,z_0}^{\pi}-z_{ref}\|,$$
and
$z_{ref}$ a given ``target'' 
state of $\mathbb{R}^n$.\\

We consider the {\em value function} ${\bf v}_k^\tau:\mathbb{R}^n\rightarrow \mathbb{R}_{\geq 0}$
defined by:
$${\bf v}_k^\tau(z_0) := \min_{\pi\in A^k}\{J_{k,\tau}(z_0,\pi)\}\equiv
\min_{\pi\in A^k}\{\|X_{k\tau,z_0}^{\pi}-z_{ref}\|\}.$$  
The function $\min$ is well-defined because the set $A$ is finite.

We consider the following {\em finite time horizon optimal control problem}: \\

Given $k\in\mathbb{N}$ and $\tau\in\mathbb{R}_{>0}$, find for each $z\in\mathbb{R}^n$
\begin{itemize}
\item the {\em value} 
${\bf v}_k^{\tau}(z_0)$, i.e.
%
$$\min_{\pi\in A^k}\{\|X_{k\tau,z_0}^{\pi}-z_{ref}\|\},$$

\item and an {\em optimal pattern}:
$$\pi^k_{\tau}(z_0) := arg\min_{\pi\in A^k}\{\|X_{k\tau,z_0}^{\pi}-z_{ref}\|\}.$$
\end{itemize}

\


We are interested here in an optimal problem
with {\em safety constraints}:
we want that all the trajectories starting in ${\cal S}$
always stay in~${\cal S}$ for $t\in [0,k\tau]$.
More precisely, we will focus on control patterns~$\pi\in A^k$ 
that are ``admissible for~$z_0\in {\cal S}$'', i.e. such that:\ 
$X_{i\tau,z_0}^{\pi}\in  {\cal S},$
for all $i\in\{1,\dots,k\}$
({\em discrete-time safety} constraint).
We will also consider a stronger admissibility criterion requiring:\ 
$X^{\pi}_{t,z_0}\in  {\cal S},$ for all $t\in [0, k\tau]$
({\em continuous-time safety} constraint).\\

In order to solve such optimal control problems, 
it is classical to {\em spatially discretize} the set ${\cal S}\subset \mathbb{R}^n$.
Given a hyper-rectangle ${\cal S}$ ,
we consider a {\em partition} of ${\cal S}$ into a finite number of 
hyper-rectangular cells. The {\em grid} ${\cal X}$ associated with ${\cal S}$
is the set of all the cell centers. 
We suppose furthermore that the radius of every cell is upper bounded 
by a given positive real $\varepsilon$:
each cell $C$ of center $x$ is such that
$\|z_0 - x\|\leq\varepsilon$, for all $z_0\in C$.
The center~$ x\in{\cal X}$ of a cell $C\subset {\cal S}$ is said to be the 
``$\varepsilon$-representative'' of all point of $C$.
Since the set of cells forms a partition of ${\cal S}$, each point $z_0\in {\cal S}$ has a unique
$\varepsilon$-representative $x\in {\cal S}$ with $\|z_0-x\|\leq \varepsilon$.\\


In this context, the idea of a Semi-Lagrangian (SL) procedure is the following: we consider the points of ${\cal X}$
as the vertices of a finite oriented graph;
there  is a connection from $ x\in {\cal X}$ 
to $ x'\in {\cal X}$ if $ x'$ is the $\varepsilon$-representative
of the Euler-based image $(x +\tau f_a( x))$ of $ x$, for some $a\in A$.
We then compute using dynamic programming
the ``path of length $k$ with minimal cost'' starting at $ x$: 
such a path is a sequence of $n+1$ 
connected points $ x\  x_k\  x_{k-1}\ \cdots\  x_1$ of
${\cal X}$ which minimises the distance $\| x_1-z_{ref}\|$.
The dynamic progamming procedure thus gives us
a {\em spatially discrete value function} ${\bf v}_k^{\tau,\varepsilon}: {\cal X}\rightarrow \mathbb{R}_{\geq 0}$,
and a {\em spatially discrete pattern function} $\pi^k_{\tau,\varepsilon}:
{\cal X}\rightarrow A^k$ which ``approximate'' 
on ${\cal S}$ their counterparts ${\bf v}_k^\tau$ and~$\pi^k_\tau$ respectively.\\

There is a vast literature on SL-schemes 
(see, e.g.,~\cite{c12,c6})
which gives numerous results to the following {\em convergence}
problem P1:\\

``Under which conditions
does the spatially-discrete value function  ${\bf v}_k^{\tau,\varepsilon}$
converge
to the value function~${\bf v}_k^{\tau}$
when $\varepsilon\rightarrow 0$ 

?''\\


Actually, when $\varepsilon$ decreases too much, 
the computations with SL-procedures become quickly impractical. We prefer 
to consider here that $\varepsilon$ is {\em fixed} (as well as $\tau$ and $k$), 
and focus on the following (local) problem P2:\\

``Given $z_0\in {\cal S}$,
under which conditions
does there exist a
pattern $\pi\in A^k$ which guarantees:
\begin{enumerate} 
\item 
the satisfaction of the safety constraint
$X_{i\tau,z_0}^{\pi}\in  S$ for all $i\in\{1,\dots,k\}$
(or $X^{\pi}_{t,z_0}\in  S$, for all $t\in [0, k\tau]$), 
\item 
while minimizing $\|X_{k\tau,z_0}^{\pi}-z_{ref}\|$ as much as possible\\
?''
\end{enumerate}

\ 

In order to solve problem P2, we use the SL-based procedure as sketched out above. Given a point $z\in {\cal S}$ of
$\varepsilon$-representative $ x\in{\cal X}$, we
apply the SL procedure to $ x$.
The procedure generates a path of the 
form $ x\  x_k\  x_{k-1}\ \cdots\  x_1$,
where $ x_k, \cdots,  x_1$
are computed using an {\em Euler scheme}, and lie by construction in ${\cal S}$.
The associated control pattern is of the form $a_k\ a_{k-1} \cdots a_1\in A^k$.
Let $\pi_i := a_k \cdots a_{i}$ for $1\leq i\leq k$.
In order, to ensure that the corresponding points
$X_{\tau,z}^{\pi_1}, X_{2\tau,z}^{\pi_2},\dots X_{k\tau,z}^{\pi_k}$ of
the {\em exact trajectory} lie also in ${\cal S}$,
we need to establish a bound on the pairwise distances:

$\| x -z_0\|$, $\| x_{k}-X_{\tau,z_0}^{\pi_1}\|$, $\| x_{k-1}-X_{2\tau,z_0}^{\pi_2}\|$, \dots,

\hspace*{\fill} $\| x_{i}-X_{(k-i+1)\tau,z_0}^{\pi_i}\|$, \dots,
$\| x_1-X_{k\tau,z_0}^{\pi_k}\|$.\\
At time $t=0$, the first distance $\| x -z_0\|$ is known to be bounded by~$\varepsilon$.
We will establish bounds $\Delta_1, \dots, \Delta_k$ on the following distances
using a recent result which gives an upper bound to 
the deviation of Euler-based trajectories 
with time (see \cite{c13}). 
%
More precisely, we 
will give an error function~$\Delta(t)$ measuring
 the distance at time $t$ 
between an approximate
(Euler-based) trajectory
starting at $ x\in {\cal X}$
given by the SL-scheme,
and  an exact trajectory starting from the cell of $x$.
In order to guarantee that the exact trajectory
always lies in the hypercube ${\cal S}$ at times 
$t=\tau,2\tau,\dots,k\tau$, we merely perform two simple
operations:
\begin{enumerate}
\item compute the ``safety margin'' of
the Euler-based trajectory, i.e., its distance
to the boundary of~${\cal S}$
at time $t=\tau, 2\tau,\dots, k\tau$, and
\item
check that this margin is always greater than the error $\Delta(t)$
at time $t=\tau, 2\tau,\dots, k\tau$. 
\end{enumerate}
The complexity of these operations is very low. 

\

\subsection{Comparison with related work}
We distinguish between works dealing with problem P1 and those dealing with P2. 
\begin{itemize}
\item Problem P1: In many papers of the 
literature on SL methods 
with state constraints (see, e.g., \cite{c1}),
the authors enforce the trajectory system to stay in ${\cal S}$
by introducing a (somehow artificial)
``penalization'' term in the cost function~$J$,
 making the cost of crossing the boundary of~${\cal S}$
prohibitive (cf. \cite{c11}).
In order to guarantee the result of convergence
of 
$v^{\tau,\varepsilon}$ 
to $v^\tau$, 
they also often make a restrictive assumption of ``controllability''.
Note however that,
in works like \cite{c7,c8,c9},
no controllability assumption is made.

In \cite{RR19} (cf. \cite{RR17}), the authors construct a sequence 
of abstractions which are more and more precise. The sequence
of value function associated with each abstraction {\em converges} to the optimal value function associated the original problem. 
The abstract transition function computes an over-approximation of the set 
of trajectories starting at neighbouring points. This over-approximation is 
computed using a growth bound 
(bounding the distance of neighboring trajectories) based on the 
Jacobian matrix of $f_a$.
More precisely, the growth bound
is a function 
mapping any ${\bf r}\in \mathbb{R}_{\geq 0}^n$ 
to $e^{Mt}{\bf r}$, where 
$M$ is a $n\times n$-matrix whose $(i,j)$-entry is $D_jf_a^i(z)$, if $i=j$
and $|D_jf_a^i(z)|$ otherwise, and $f_a^i(z)$ denotes the $i$-th component
of vector $f_a(z)$.
By comparison,
our work here can be seen as a particular case of \cite{RR19} where one 
uses, for each of the $n$ components, a uniform growth bound,
mapping
$r\in\mathbb{R}_{\geq 0}$ to $e^{\lambda_a t}r$.
The counterpart of the convergence result of \cite{RR19} for the value function,
would state in our context that the synthesized control converge towards the 
optimal control as 
$\varepsilon$ tends to 0.
However, this does not seem true (unless adding very restrictive assumptions), which leads us to focus on problem P2 instead 
of P1.

\item Problem P2: In the work of \cite{c4,c5},
%
the authors pursue an objective similar to ours: providing
a (finite time-horizon) optimal control procedure with 
a formal guarantee of constraint satisfaction (safety). However
they do not use SL-schemes, but 
perform a reachability analysis
based on over-approximative state set
representations ({\em zonotopes}, cf. \cite{c15,c16}).

In \cite{c3},
the authors also provide a formal guarantee
of safety property. Contrarily to \cite{c4,c5}, 
they do use SL-schemes. They also
focus to (periodically) sampled systems as we do.
%
However, they 
still
perform a form of reachability analysis
similar to~\cite{c4,c5}, using
convex polytopes as
state set
representations. Their growth bound are not based on
OSL constants as here, but rather on
overapproximations of Lagrange remainders in Taylor series.
%
\end{itemize}

\section{Sufficient Conditions for Reachability with Safety}\label{sec:SL}


Given  a starting point $z_0\in {\cal S}$ and a mode $a\in A$, we denote 
by  $\tilde{X}_{\tau,y}^a$ the Euler-based image of $z_0$ at time $t=\tau$ via $a$.
We have:
$$\tilde{X}_{\tau,z_0}^a := z_0+ \tau\ f_a(z_0).$$

\ 

The {\em set of admissible modes for~$ x\in {\cal X}$} is defined by:
$$A_{\tau}( x) := \{a\in A\ |\ \tilde{X}_{\tau, x}^a\in  {\cal S}\}.$$

\

The function $next^{a}: {\cal X}\rightarrow {\cal X}\cup \{\bot\}$ 
is defined by:

\begin{itemize}
\item if $a\in A_\tau( x)$, then: 
    $next^{a}( x)= x'$, where
    $ x'$ 
    is the $\varepsilon$-representative $\tilde{X}_{\tau, x}^{a}$ ,
\item otherwise (i.e., $\tilde{X}_{\tau, x}^{a}\not\in  S$): $next^{a}( x)=\bot$.
\end{itemize}

\vspace*{2mm}

\ 

For a pattern $\pi\in A^k$, the function~$next^\pi:{\cal X}\rightarrow{\cal X}\cup\{\bot\}$ is defined as follows:

\begin{itemize}
\item if $\pi=a$ for some $a\in A$, then $next^\pi( x)=next^a( x)$,
\item if $\pi$ is of the form $a\cdot \pi'$; 

\begin{itemize}
\item if $next^a( x)\neq \bot$, then \\
$next^\pi( x)=next^{\pi'}(next^a( x))$,
\item otherwise, $next^\pi( x)=\bot$.
\end{itemize}
\end{itemize}

\vspace*{2mm}

\ 

It is easy to show, using the definition of $next$:\\

\begin{proposition}\label{prop:1}
Let~$x\in{\cal X}$, and 
$\pi_k\in A^k$ a pattern of the form $a_k\ a_{k-1}\cdots\ a_1$.
Let  us write $\pi_i := a_k\cdots a_i$ for $1\leq i\leq k$,
and $x_{k+1} := x$.

If $next^{\pi_k}(x)\in{\cal X}$, then 
there exists a sequence of points 
of the form $x_{k+1} x_k \cdots x_1\in {\cal X}^{n+1}$
with,
for~all~$1\leq i\leq k$:
\begin{itemize}
\item $\tilde{X}_{\tau, x_{i+1}}^{a_i}\in {\cal S}$, 
\item $x_i = next^{a_i}(x_{i+1})=next^{\pi_i}(x)$, and
\item
$\|x_i - \tilde{X}_{\tau, x_{i+1}}^{a_i}\|\leq \varepsilon$.
\end{itemize}
\end{proposition}

\

\begin{definition}
For all point $ x\in {\cal X}$, the {\em spatially discrete
value function} ${\bf v}^{\tau,\varepsilon}_k:{\cal X}\rightarrow \mathbb{R}_{\geq 0}\cup\{\infty\}$ 
is defined by:
\vspace*{2mm}
\begin{itemize}
\item for $k=0$, ${\bf v}_k^{\tau,\varepsilon}( x)=\| x\|$,
\item for $k\geq 1$,
    \begin{itemize}
    \item if 
    $A_{\tau}( x)=\emptyset$:\ ${\bf v}_{k}^{\tau,\varepsilon}( x)=\infty$,

     \item  if $A_{\tau}( x)\neq \emptyset$:\\
    ${\bf v}_{k}^{\tau,\varepsilon}( x)=\min_{a\in A_\tau( x)}\{{\bf v}_{k-1}^{\tau,\varepsilon}(next^a( x))\}$.
\end{itemize}
\end{itemize}
\end{definition}

\

If ${\bf v}_k^{\tau,\varepsilon}( x)\neq \infty$, 
one defines the {\em approximate  optimal pattern of length $k$}
associated
to $ x$,
denoted by $\pi_k^{\tau,\varepsilon}( x)\in A^k$,  recursively by:\\

\begin{itemize}
\item if $k=0$, $\pi_k^{\tau,\varepsilon}( x)=\mbox{nil}$,\\

\item if $k\geq 1$, $\pi_k^{\tau,\varepsilon}( x) = {\bf a}_k( x) \cdot \pi'$
where
$${\bf a}_{k}( x)=arg \min_{a\in A_\tau( x)}\{{\bf v}_{k-1}^{\tau,\varepsilon}(next^a( x))\}$$
and $\pi' =\pi_{k-1}^{\tau,\varepsilon}( x')$
\ \ with \ $ x'=next^{{\bf a}_k( x)}( x)$.
\end{itemize}

\

Using the value function
${\bf v}_k^{\tau,\varepsilon}$ 
it is thus easy to
construct
an SL procedure $PROC_k^{\tau,\varepsilon}$ which takes a point $ x\in {\cal X}$ as input, and returns, in case of success
(i.e., when ${\bf v}_k^{\tau,\varepsilon}( x) \geq 0$), a pattern
$\pi_k^{\tau,\varepsilon}\in A^k$ with $next^{\pi_k^{\tau,\varepsilon}}(x)\in{\cal X}$.
We now define, for such a pattern $\pi_k^{\tau,\varepsilon}$ output by 
$PROC_k^{\tau,\varepsilon}(x)$,
a value $\Delta(\pi_k^{\tau,\varepsilon})$ which gives us an {\em upperbound}
to~$\|X_{k\tau,z_0}^{\pi_k^{\tau,\varepsilon}}-next^{\pi_k^{\tau,\varepsilon}}( x)\|$, for any~$z_0\in B( x,\varepsilon)$ (i.e., any $z_0$ such that
: $\|z_0 - x\|\leq \varepsilon$).\\

\begin{definition}\label{def:delta}
Let $\mu$ be a given positive constant. Let us define, for all 
$a\in A$ and $t\in [0,\tau]$,
$\delta^a_{t,\mu}$ as follows:
\begin{itemize}
\item  if $\lambda_a <0$:
{\small{
$$\delta^a_{t,\mu}=\left(\mu^2 e^{\lambda_a t}+
 \frac{C_a^2}{\lambda_a^2}\left(t^2+\frac{2 t}{\lambda_a}+\frac{2}{\lambda_a^2}\left(1- e^{\lambda_a t} \right)\right)\right)^{\frac{1}{2}}$$
}}

\item if $\lambda_a = 0:$
{\small{
$$\delta^a_{t,\mu}= \left( \mu^2 e^{t} + C_a^2 (- t^2 - 2t + 2 (e^t - 1)) \right)^{\frac{1}{2}}$$
}
}


\item if $\lambda_a > 0:$
{\small{
\begin{multline*}
\delta^a_{t,\mu}=\left(\mu^2 e^{3\lambda_a t}+\right.
\\
\left.
\frac{C_a^2}{3\lambda_a^2}\left(-t^2-\frac{2t}{3\lambda_a}+\frac{2}{9\lambda_a^2}
\left(e^{3\lambda_a t}-1\right)\right)\right)^{\frac{1}{2}}
\end{multline*}
}}
\end{itemize}
where $C_a$ and $\lambda_a$ are constants defined in Section~\ref{ss:defs}.
\end{definition}

\ \\

\begin{proposition}\label{prop:formats17} \cite{c13}
Given $ x\in \mathbb{R}^n$, we have,
for all $a\in A$
and all $z_0\in B(x, \varepsilon)$ (i.e., $z_0$ such that
$\|z_0-x\|\leq \varepsilon$):
$$\|X_{\tau,z_0}^{a}-\tilde{X}^{a}_{\tau,  x}\|
\leq \delta^{a}_{\tau,\varepsilon}.$$
\end{proposition}

\ 

\begin{definition}\label{def:Delta}
Let us define 
$\Delta(a_k\cdots a_1)$ recursively by:
\vspace*{2mm}
\begin{itemize}
    \item $\Delta(a_i)=\delta^{a_i}_{\tau,\varepsilon}$ for $i=1$, and
    \item $\Delta(a_i\cdots a_1)=\delta^{a_i}_{\tau,\mu}$ 
with $\mu = \varepsilon+\Delta(a_{i-1}\cdots a_{1})$, for $i\geq 2$.
    \end{itemize}
\end{definition}

\ 

In the rest of the paper, we suppose that $k\in \mathbb{N}$
and~$\tau, \varepsilon\in\mathbb{R}_{>0}$ are given and fixed. 
So, for the sake of notation simplicity,
we will abbreviate ${\bf v}_k^{\tau,\varepsilon}$
as ${\bf v}_k$. We abbreviate similarly
$\pi_k^{\tau,\varepsilon}$ and $PROC_k^{\tau,\varepsilon}$
as $\pi_k$ and  $PROC_k$ respectively.
We will suppose also that we are given a compact set ${\cal S}\subset \mathbb{R}^n$
as well as a ``target'' set $R\subset {\cal S}$.\footnote{We suppose implicitly that $R$ contains the target point $z_{ref}$, so $R$ can be seen as a neighborhood of $z_{ref}$.}
We have:\\

\begin{lemma}\label{lemma0}
Let $ x\in{\cal X}$
and $\pi_k\equiv a_k\cdots a_1\in A^k$ the
pattern generated by $PROC_k( x)$ with 
$next^{\pi_k}(x)\in{\cal X}$.
We have, 
for all~$z_0\in B( x,\varepsilon)$: 
\begin{enumerate}
\item $\|X_{k\tau,z_0}^{\pi_k}-\tilde{X}^{a_1}_{\tau, x_{2}}\| \leq \Delta(\pi_k)$, 

with $x_2 := next^{a_k\cdots a_2}(x)$ for $k\geq 2$, and $x_2 := x$ for $k=1$; 
%
\item
$\|X_{k\tau,z_0}^{\pi_k}- next^{\pi_k}(x) \|\leq \Delta(\pi_k)+\varepsilon.$
%
\end{enumerate}
\end{lemma}
%
%

\vspace{1em}
\begin{proof}
Let us prove items 1-2 by induction on $k$.

Let us first prove item 1 for the base case $k=1$. We have $\pi_k=\pi_1=a_1$ and $x_2=x$.
We have to prove, when\\
$\|x -z_0\|\leq \varepsilon$:
\ \ $\|X_{\tau,z_0}^{a_1}-\tilde{X}^{a_1}_{\tau, x}\| \leq \Delta(a_1)=\delta_{\tau,\varepsilon}^{a_1}$. This inequation holds by Proposition~\ref{prop:formats17}, 
and the proof of item~1 is done.
The proof of item~2 of the base case follows
from item~1
and Proposition~\ref{prop:1},
using triangular inequality.

Let us now consider the induction step.
We have to prove the following induction conclusion:
\begin{enumerate}
\item $\|X_{(k+1)\tau,z_0}^{a_{k+1}\cdots a_1}-\tilde{X}^{a_1}_{\tau, x_{2}}\| \leq \Delta(a_{k+1}\cdots a_1)$, 

with $x_2 := next^{a_k\cdots a_2}(x)$ 
%
\item
$\|X_{(k+1)\tau,z_0}^{a_{k+1}\cdots a_1}- next^{a_{k+1}\cdots a_1}(x) \|$

\hspace*{\fill}$\leq \Delta(a_{k+1}\cdots a_1)+\varepsilon$.
\end{enumerate}

We have by induction hypothesis:
\begin{enumerate}
\item $\|X_{k\tau,z_0}^{a_{k+1}\cdots a_2}-\tilde{X}^{a_2}_{\tau,x_3}\|\leq \Delta(a_{k+1}\cdots a_2)$, 

with $x_3 := next^{a_{k+1}\cdots a_2}(x)$,
and

\item $\|X_{k\tau,z_0}^{a_{k+1}\cdots a_2}-next^{a_{k+1}\cdots a_2}(x)\|\leq \mu$,\\
\hspace*{\fill} with $\mu := \Delta(a_{k+1}\cdots a_2)+\varepsilon$.
\end{enumerate}
Besides, by Definition~\ref{def:Delta}: $\Delta(a_{k+1}\cdots a_1)=\delta_{\tau,\mu}^{a_1}$.\\
Applying Proposition~\ref{prop:formats17},
with $z_2=X^{a_{k+1}\cdots a_2}_{k\tau,z_0}$ and\\
$x_2 = next^{a_{k+1}\cdots a_2}(x)$, we have:
$$\|X_{\tau,z_2}^{a_1}-\tilde{X}^{a_1}_{\tau,x_2}\|\leq \delta^{a_1}_{\tau,\mu}$$
since $\|x_2-z_2\|\leq\mu$ by item 2 of induction hypothesis.
It follows
%
$$\|X^{a_{k+1}\cdots a_1}_{(k+1)\tau,z_0}-\tilde{X}^{a_1}_{\tau, x_2}\|\leq \delta^{a_1}_{\tau,\mu} = \Delta(a_{k+1}\cdots a_1).$$
This achieves the proof of the item 1 of the induction conclusion.
The item~2 of the induction conclusion then follows
from item~1
and Proposition~\ref{prop:1},
using triangular inequality.
This completes the proof of the induction step.
\end{proof}



Using item 2 of Lemma~\ref{lemma0}, it is easy to show:\\

\begin{theorem}\label{th:MINIMATOR} (sufficient conditions of safety and $k$-reachability)
Let $ x\in {\cal X}$, 
and~${\bf \pi}_k\equiv a_k\cdots a_1\in A^k$ the pattern generated by~$PROC_k(x)$ with 
$next^{\pi_k}(x)\in{\cal X}$.
Suppose, for all~$1\leq i\leq k$:\\

\begin{itemize}
    \item $(H_1^i)$:\ \ \ $B(next^{\pi_i}( x), \Delta(\pi_i)+\varepsilon)\subseteq  {\cal S}$, and 
    \item $(H_2^k)$:\ \ \ $B(next^{\pi_k}( x), \Delta(\pi_k)+\varepsilon)\subseteq  R$,
    \end{itemize}
\vspace*{2mm}

where~$\pi_i := a_k \cdots a_{i}$. 
Then we have, for
all~$z_0\in B( x,\varepsilon)$\footnote{In particular, for any $z_0\in  S$ of $\varepsilon$-representative~$ x$.}\\

    \begin{itemize}
        \item 
    $X_{(k-i+1)\tau,z_0}^{\pi_i}\in  S$ for all $1\leq i\leq k$ \\
\hspace*{\fill} ({\em discrete-time safety}), 

and 
\item $X_{k\tau,z_0}^{\pi_k}\in  R$ \ \ \ \ \ \ \ \ \hspace*{\fill} ($k$-{\em reachability}).
    \end{itemize}
    Furthermore, assuming that, for all $a\in A$, 
$\delta_{t,\varepsilon}^a$ is a convex function
for $t\in [0,\tau]$
(i.e., $\frac{d^2 (\delta^a_t)}{dt^2}>0$ for all $t\in [0,\tau]$
\footnote{The sign of $\frac{d^2 (\delta^a_t)}{dt^2}$ on $[0,\tau]$
depends on the value of the constants $C_a$ and $\lambda_a$
occurring in Definition~\ref{def:delta}; knowing these constant
values, the sign is easy to determine (see \cite{c13}).}),
we have: 

$X_{t,z_0}^{\pi_k}\in  S$ for all $t\in [0,k\tau]$ \ \ \ \hspace*{\fill}({\em dense-time safety}).
\end{theorem}
\vspace*{2mm}

\ 

Suppose in particular that conditions $(H_1^i)$-$(H_2^k)$ hold for a set of points~${\cal Y}\subseteq {\cal X}$ which {\em $\varepsilon$-covers} $R$, i.e., such that:
$R\subseteq \bigcup_{ x\in{\cal Y}} B( x,\varepsilon)$. 
In this case, the procedure $PROC_k$ gives  us a guarantee of ``$(R,S)$-stability'' as defined in~\cite{c14}. 
By Theorem~\ref{th:MINIMATOR}, we know indeed that, for all $z_0\in R$ of representative $ x\in {\cal X}$, the pattern
$\pi_k$ generated by $PROC_k(x)$ applied to~$z_0$ yields a trajectory that reaches at $t=k\tau$ a point~$z'$ of~$R$  (while always staying in ${\cal S}$
for $0\leq t\leq k\tau$); the 
process can then be iterated to $z'$, and so on repeatedly.
This means that, via the set of patterns $\pi_k$ associated to 
elements of ${\cal Y}$, one can control any trajectory starting at $R$
in order to make it return to $R$ periodically every $k\tau$ seconds, 
and stay in ${\cal S}$ for all $t\geq 0$ (``$(R,S)$-stability'').\footnote{$R$ can be seen as a special case of {\em viability kernel} for ${\cal S}$ 
(see, e.g., \cite{c10}) since
any trajectory starting from~$R$ can be controlled in order to stay inside~${\cal S}$ forever.}

The SL-based procedure $PROC_k$ can thus replace advantageously the brute-force enumeration strategy implemented in tool MINIMATOR \cite{c17}: the time complexity of MINIMATOR procedure is indeed
$O(m^k\times N)$ where $m$ is the number of modes, $N$ the number of cells
and $k$ the time-horizon length, while the complexity of  $PROC_k$ is  
$O(m\times k\times N)$.\\

\subsection{Description of the implementation}

The procedure is implemented in Octave. It is composed of 9 functions and a main script totalling 500 lines of code. 
For comparison, the tool MINIMATOR uses 28 functions for a total of 2000 lines of code. 

The computations are realised in a virtual machine running Ubuntu 18.06 LTS, having access to one core of a 2.3GHz Intel Core i5, associated to 3.5GB of RAM memory. 

Note that the accuracy of the Euler approximation can be optionally increased by using a smaller time step.  
The time-step $h$ used for Euler approximation is not necessarily equal
 to the control sampling period, but is in general
a {\em submultiple} of $\tau$ ($\tau= p\times h$ where $p$ is a natural number
greater than 1).
\\


%

\begin{example}\label{ex:temp_opt} (2-tanks)

In this example, we illustrate the approach given above for $(R,S)$-stability on a two tank example. 
The two-tank  system  is a linear example taken from \cite{c18}. 
 The system consists of two tanks and two valves.
 The first valve adds to the inflow of tank 1 and the second valve is a drain valve for tank 2. 
 There is also a constant outflow from tank 2 caused by a pump. The system is linearized at a desired
 operating point. The objective is to keep the water level in both tanks 
 within limits using a discrete open/close switching strategy for the valves. 
 Let the water level of tanks 1 and 2 be given by $x_1$ and $x_2$ respectively. 
 The behavior of $x_1$ is given by $\dot x_1 = -x_1 - 2$ when the tank 1 valve is closed, 
 and $\dot x_1 = -x_1 + 3$ when it is open. Likewise,
 $x_2$ is driven by $\dot x_2 = x_1$ when the tank 2 valve is closed and $\dot x_2 = x_1 - x_2 - 5$ when it 
 is open.

Let $S=[-2,3]\times[-1,2]$, $R=[-1.5,2.5]\times[-0.5,1.5]$, $N = 10 \times 10$ the number of cells, $\varepsilon=0.33$,
$\tau=0.1$.
The proof of $(R,S)$-stability is obtained for $k=5$, it takes 7.34 seconds. By comparison,
MINIMATOR takes 25.53 seconds to obtain a controller without any optimality result. 
Simulations of the $(R,S)$-stability  controller are given in Figure~\ref{fig:two_tank}.

\begin{figure}
\includegraphics[width=0.45\textwidth, clip, trim = 1cm 6cm 1cm 7cm]{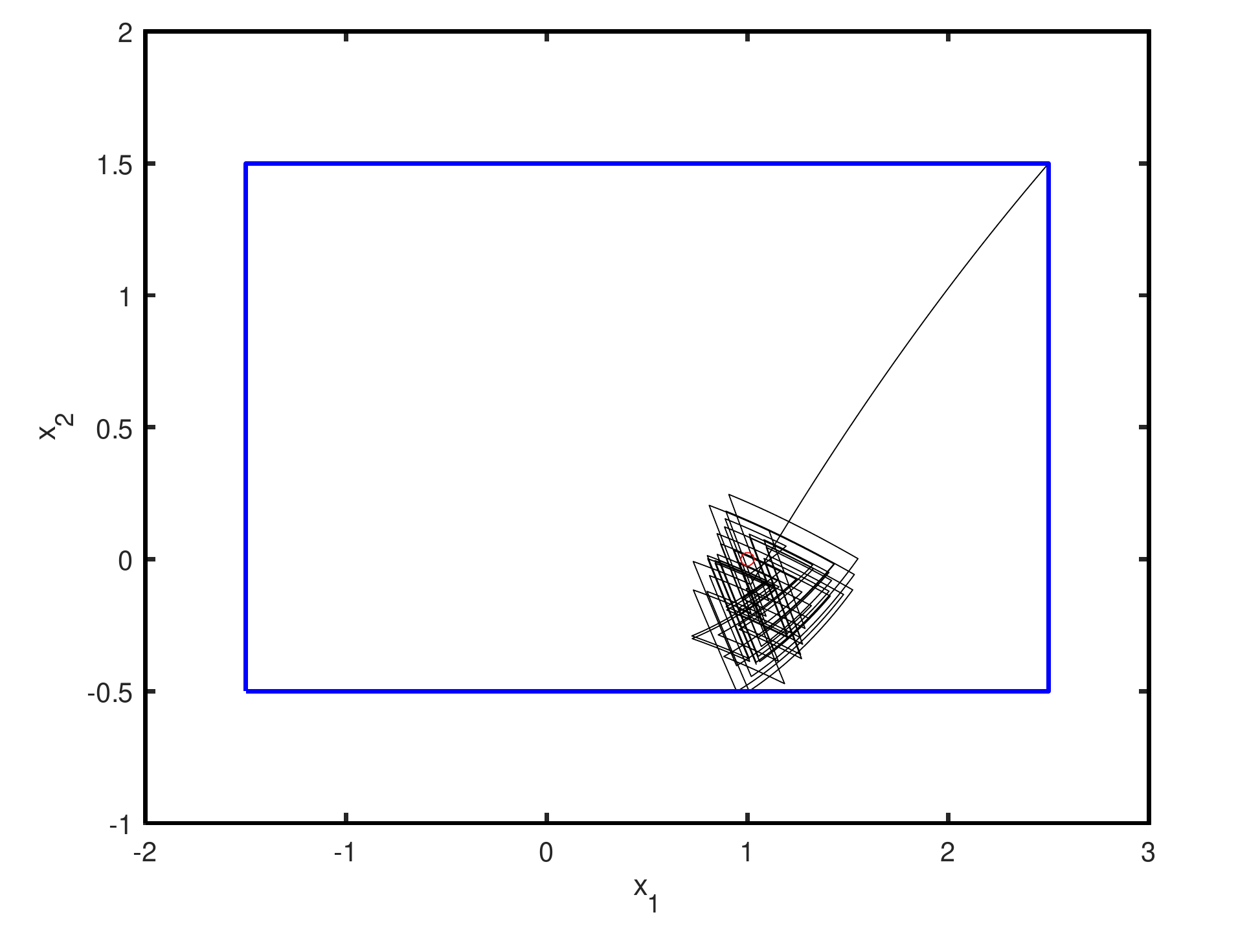}
\includegraphics[width=0.45\textwidth, clip, trim = 1cm 6cm 1cm 7cm]{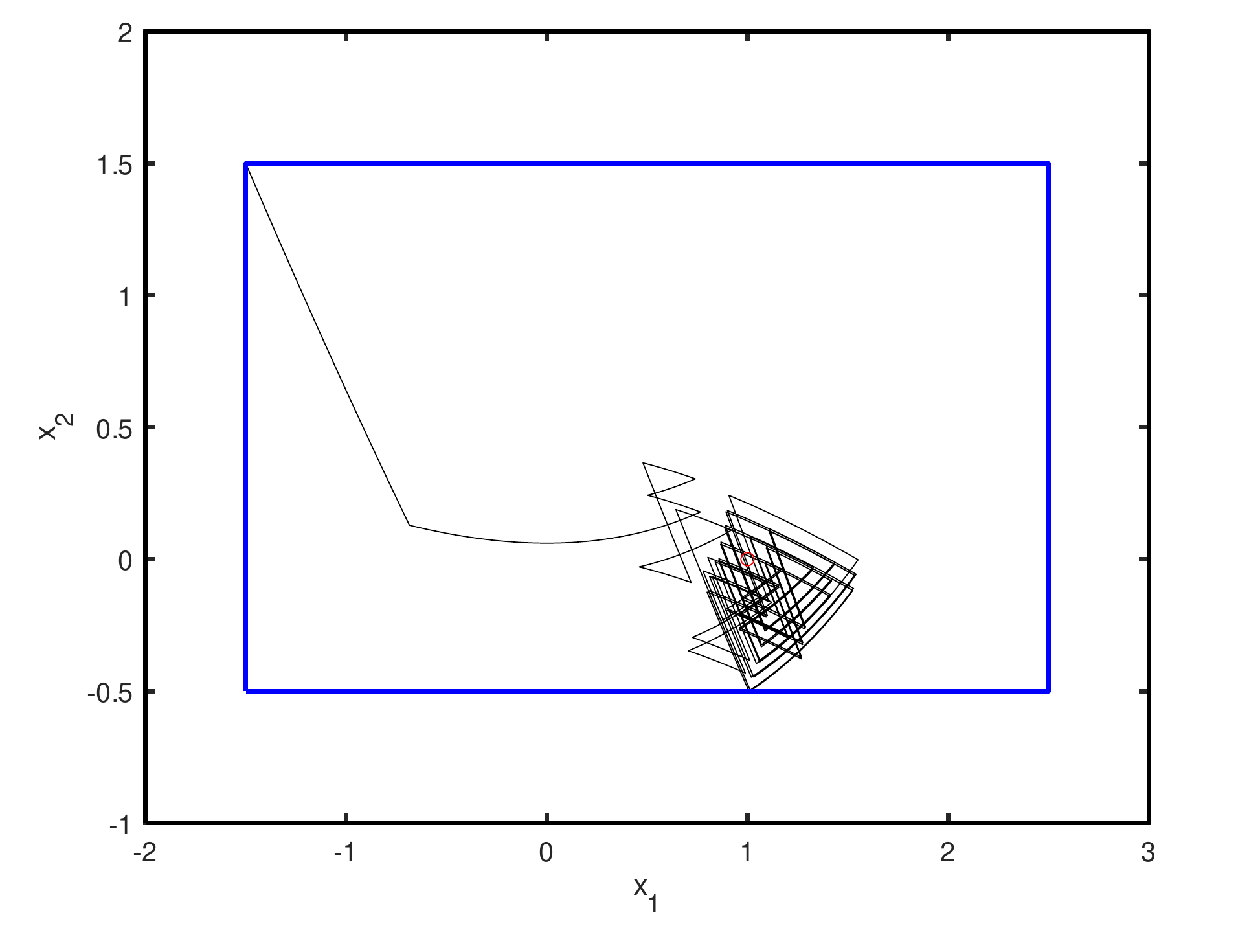}
\caption{Simulations of the $(R,S)$-stability controller on the two tank example. The safety set is $S=[-2,3]\times[-1,2]$, the recurrence set $R=[-1.5,2.5]\times[-0.5,1.5]$, the blue box is the set $R$, the red circle is the objective here chosen as $(1.0,0.0)$. The trajectory of the system is in black for two initial conditions: $(2.5,2.5)$ (top) and $(-1.5,1.5)$ (bottom).}
\label{fig:two_tank}
\end{figure}

\end{example}

\section{Extensions}

We now explain how to extend the method to stochastic ODEs and differential games.

\subsection{Stochastic switched systems}

Let us consider a stochastic switched system defined by
\begin{equation} \d X_t = f_a(X_t)\d t+g_a(X_t)\d W_t,
\quad \quad X_0 = x_0,
\label{eq:sde}
\end{equation}
where  $W_t$ is a standard $m-dimensional$ Brownian motion, and suppose that 
for all $a \in A$:

(H1) $f_a : \mathbb{R}^n\rightarrow \mathbb{R}^n$ is a continuously differentiable function
whose derivative grows at most polynomially,

(H2) $g_a=(g_{a_{i,j}})_{i\in\{1,\dots,n\},j\in\{1,\dots,m\}}: \mathbb{R}^n\rightarrow\mathbb{R}^{n\times m}$ is a globally Lipschitz 
continuous function,

 (H3) $f_a$ is globally one-sided Lipschitz.\\

Under the above-mentioned hypotheses, we can establish 
bounds  $\delta^a_{t,\varepsilon}$ similar to Definition \ref{def:delta}
for stochastic switched systems using the {\em tamed Euler scheme} \cite{HMS02}.
The detail of this bound is given in Appendix for a single mode stochastic switched system ({\em i.e.} a stochastic differential 
equation).
We refer the reader to \cite{tamed} for the details of the error bounding for stochastic switched systems.
 The result is stated as follows for a single switching step integration:\\
 
\begin{proposition}[\cite{tamed}]\label{prop:stochastic}
Consider two points $x_0$ and $z$ in $\mathbb{R}^n$,and
a positive real number $\varepsilon$.
Suppose that $x_0\in B(z,\varepsilon)$.
Let us denote by $\tilde{X}_{t,z}$ the tamed Euler approximation of $X_t$ starting from initial point $z$ in \eqref{eq:sde}.
Then $\EE X_{t,x_0}\in B(\tilde{X}_{t,z},\delta^a_{t,\varepsilon})$
for all $t\in[0,\tau]$, where $\EE$ is the symbol of expected value.
\end{proposition}

\vspace{1em}

\begin{example}\label{ex:stochastic} (Stochastic system)
Consider the system (see (\cite{ZEMAL14,ZAG15})):

$dx_1 = (-0.25 x_1 + u x_2 +(-1)^u 0.25)dt + 0.01 x_1 dW_{t}^1$

$dx_2 = ((u-3) x_1  - 0.25 x_2 + (-1)^u (3-u)) dt + 0.01 x_2 dW_{t}^2$\\
where $u=1,2$. \\

%

%
%

%
We can apply the above procedure $PROC^{\tau,\varepsilon}_k$ in order to
minimize the average distance of the state to the origin after  a  given number of steps. 
We consider a  switching period $\tau=0.5$ subdivided in time steps
of size  $\Delta_t=10^{-4}$.
Consider the interest set $R=B((0,0),\rho)$
with $\rho = 7$,
discretized with an accuracy $\varepsilon=0.57$. 
We compute (sub)optimal patterns for the entire set, using different lengths of patterns, and simulate the induced controller for 200 initial conditions randomly selected in $R$. 
Simulations are given in Figure \ref{fig:optimal_ex2}.
The procedure took $11.8$ seconds of computation for patterns 
of length $1$, $47.4$ seconds of computation for patterns 
of length $3$.

\begin{figure}[ht]
\centering
\begin{tabular}{c}
\includegraphics[width=0.4\textwidth]{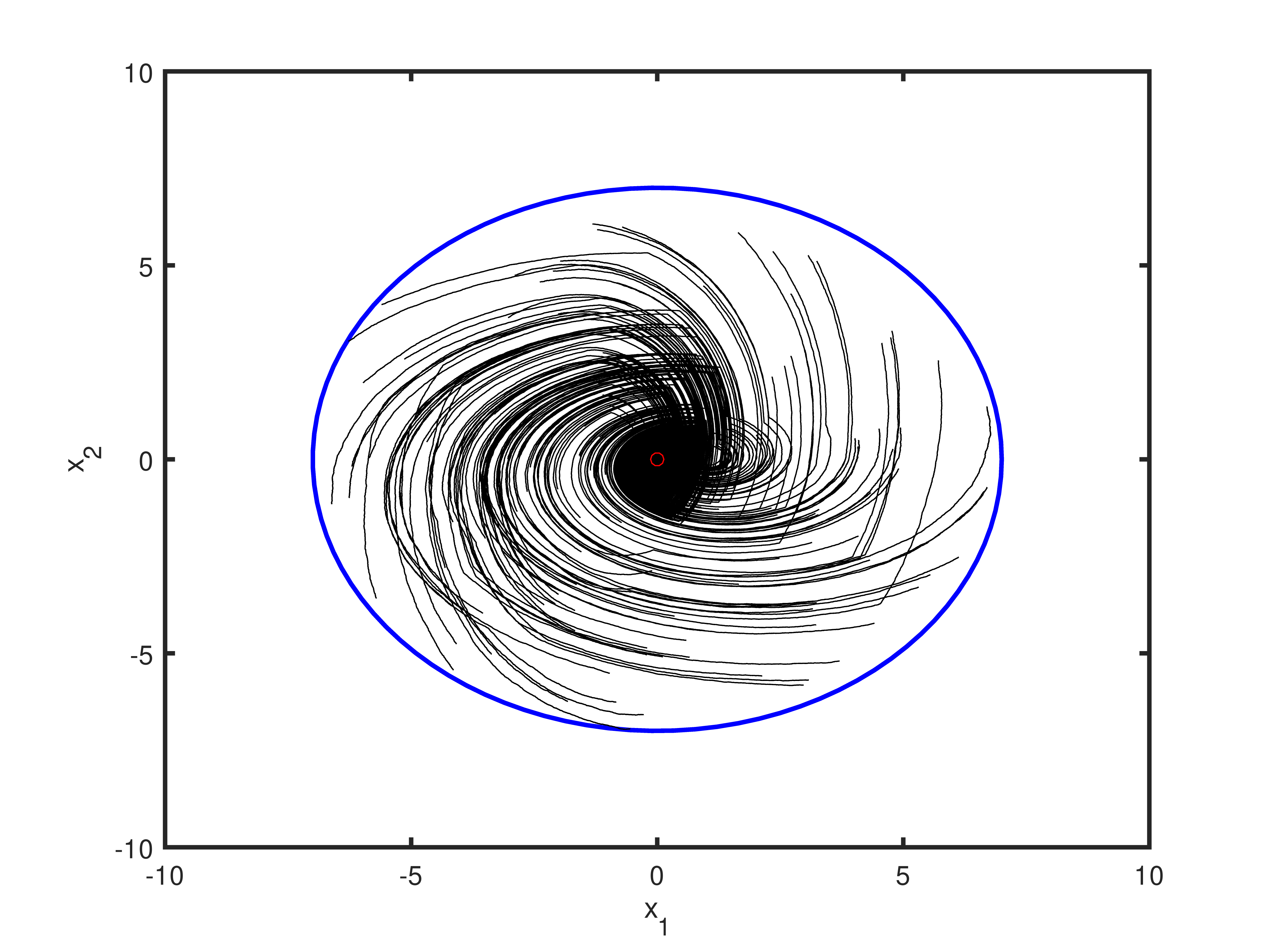}
\\
\includegraphics[width=0.4\textwidth]{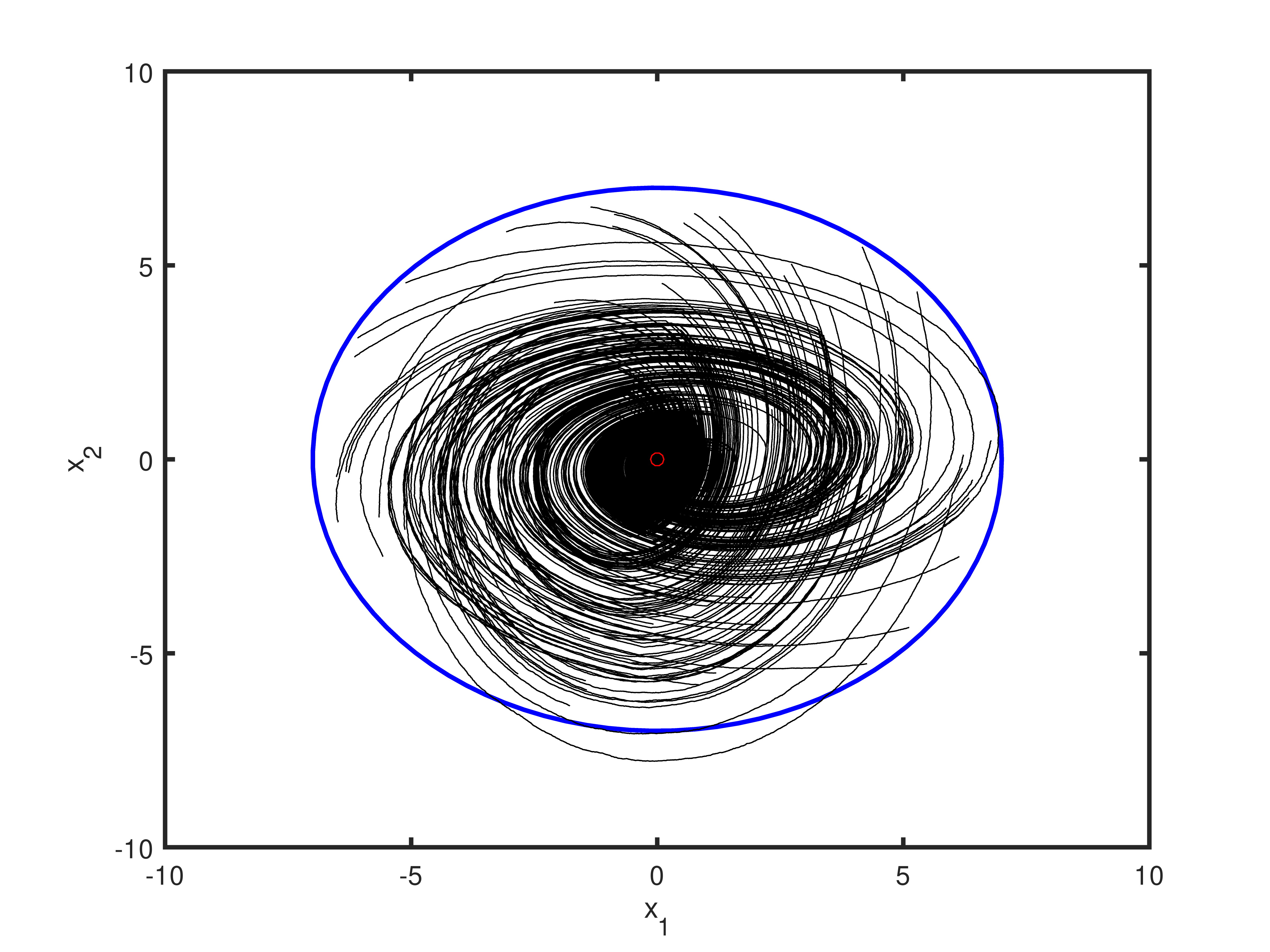}
\end{tabular}

\caption{Simulations of Example \ref{ex:stochastic} 
with the controller induced by $PROC_k$, for 
patterns of length $1$ (top), length $3$ (bottom).
The blue circle is the set $R=B((0,0),7)$, the red marker is the target state (the origin), the black lines are the controlled trajectories.}
\label{fig:optimal_ex2}
\end{figure}

 \end{example}

\subsection{Sampled Pursuit-Evasion Games with Safety}

Let us explain here how one can extend our SL-based method to Pursuit-Evasion games,
closely following the work of \cite{c12}. The ODEs are defined by:

$$(\dot{z}_1(t),\dot{z}_2(t))=(f_a^1(z_1(t)),f_b^2(z_2(t))),\ \ \ \ \ t>0$$
$$(z_1(0),z_2(0))=(z_1^0,z_2^0)$$
where $(z_1(t),z_2(t))\in \mathbb{R}^{n}\times \mathbb{R}^{n}$,
$(z_1^0,z_2^0)\in \mathbb{R}^{n}\times \mathbb{R}^{n}$, $a\in A$, $b\in B$,
and
$f^1_a$ and $f^2_b$ are functions of $\mathbb{R}^n\rightarrow \mathbb{R}^n$.
Given a safety set ${\cal S}=S_1\times S_2\subset \mathbb{R}^n\times \mathbb{R}^{n}$, the (dense-time) safety constraint is:
$z_1(t)\in  S_1$, $z_2(t)\in  S_2$.
The target set is defined by:

$R=\{(z_1,z_2)\in\mathbb{R}^{n}\times \mathbb{R}^n:\ \|z_1-z_2\|\leq \gamma\}$,\ \ $\gamma\geq 0$\\

\begin{example}\label{ex:2} (The Tag-Chase game with constraints) \cite{c12}

Two players $1$ (pursuer) and $2$ (evader) which run one after the other in the same 2-dimensional domain (courtyard), so that the game is set in ${\cal S}=S_1^2\subset \mathbb{R}^4$.
Players $1$ and $2$ can run in every direction with velocity $V_1$ and $V_2$ respectively.
The control sets are of the form
$$A=\{\alpha_1,\dots,\alpha_{m_1}\},\ \ 
B=\{\beta_1,\dots,\beta_{m_2}\}.$$
We have the dynamics for $z_1=(i_1,j_1)$ and $z_2=(i_2,j_2)$:\\

1:\ \ \ $\dot{i}_1=V_1\ sin\alpha$\ \  ;\ \  $\dot{j}_1=V_1\ cos\alpha$

2: \ \ $\dot{i}_2=V_2\ sin\beta$\ \  ;\ \  $\dot{j}_2=V_2\ cos\beta$\\
\\
where $\alpha\in A$ is the direction for $1$, and $\beta\in B$ is the direction for $2$ ($\alpha$ and $\beta$ are the angles between the~$j$-axis and the velocities for $1$ and $2$).
%
For $z_1\equiv (i_1,j_1)$
and $z_2\equiv (i_2,j_2)$, the capture occurs when $z\equiv (z_1,z_2)\in R\equiv R_1\times R_2$ with
$$R=\{((i_1,j_1),(i_2,j_2))\in {\cal S}: \sqrt{(i_1-i_2)^2+(j_1-j_2)^2}\leq \gamma\}.$$
\end{example}

We build a partition of ${\cal S}=S_1\times S_2$ and construct
a grid ${\cal X}={\cal X}_1\times{\cal X}_2$
by performing separately the operations described in Section~\ref{sec:SL}
on ${\cal S}_1$ and ${\cal S}_2$.\\

\begin{definition} 
For~$x=(x_1,x_2)\in {\cal X}$, the {\em sets 
of admissible controls} 
$A_{\tau}(x_1)$ and $B_{\tau}(x_2)$
w.r.t. ${\cal S}_1$ and ${\cal S}_2$ respectively, are defined  by:
$$A_{\tau}(x_1)=\{a\in A:\ x_1+\tau f^1_a(x_1)\in   S_1\},$$
$$B_{\tau}(x_2)=\{b\in B:\ x_2+\tau f^2_b(x_2)\in   S_2\}.$$
%

\end{definition}

\vspace*{2mm}
\

Let denote by~$next^{a,b}: {\cal X}_1\times {\cal X}_2\rightarrow {\cal X}_1\times  {\cal X}_2$, the function defined, for $x=(x_1,x_2)$ by:
$$next^{a,b}(x) := (next^a(x_1),next^b(x_2)),$$
where $next^a:{\cal X}_1\rightarrow {\cal X}_1$ and 
$next^b: {\cal X}_2\rightarrow {\cal X}_2$
are defined as in Section~\ref{sec:SL}.\\

\begin{definition}
The {\em value function} ${\bf v}_k:{\cal X}\rightarrow \mathbb{R}_{\geq 0}\cup \{\infty\}$ 
is defined, for all $x\equiv (x_1,x_2)\in{\cal X}$
with $x_1\equiv (i_1,j_1)$ and $x_2\equiv (i_2,j_2)$, by:
\begin{itemize}
\item For $k=0$, ${\bf v}_k(x)=\sqrt{(i_1-i_2)^2+(j_1-j_2)^2}$;
%
\item For $k\geq 1$,
\begin{itemize}
\item if $A_{\tau}(x_1)=\emptyset\ \vee\ B_\tau(x_2)=\emptyset$, 
${\bf v}_{k}(x)=\infty$;
%
\item
if $A_{\tau}(x_1)\neq \emptyset\ \wedge\ B_{\tau}(x_2)\neq \emptyset$,
$${\bf v}_{k}(x) =\max_{b\in B_\tau(x_2)}\min_{a\in A_\tau(x_1)}\{{\bf v}_{k-1}(next^{a,b}(x))\}.$$
\end{itemize}
\end{itemize}
\end{definition}
%

\


Similarly to what has been done in Section~\ref{sec:SL}, one can construct,
for $x=(x_1,x_2) \in{\cal X}$ with ${\bf v}_k(x)\geq 0$,
a procedure $PROC_k$ which returns a pattern $(\pi_k^1,\pi_k^2)\in A^k\times B^k$
with $next^{\pi_k^1}(x_1)\in{\cal X}_1$ and $next^{\pi_k^2}(x_2)\in{\cal X}_2$.\footnote{Note that the 1st element of the sequence~$(\pi_k^1,\pi_k^2)\in A^k\times B^k$ is of the form~$({\bf a}_k, {\bf b}_k) = arg \max_{b\in B_\tau(x_2)}\min_{a\in A_\tau(x_1)} \{{\bf v}_{k-1}(next^{a,b}(x))\}$.}
The counterpart of Theorem~\ref{th:MINIMATOR} is:\\

\begin{theorem} (sufficient conditions of safety and $k$-capture)
\label{th:capture}
Consider a point  $x=(x_1,x_2)\in {\cal X}$,
and let $(\pi_k^1,\pi_k^2)\equiv (a_k\cdots a_1, b_k\cdots b_1)$ be the
pattern generated by $PROC_k(x)$ with
$next^{\pi_k^1}(x_1)\in{\cal X}_1$ and $next^{\pi_k^2}(x_2)\in{\cal X}_2$.
Suppose that, for all~$1\leq i\leq k$:

\begin{enumerate}
\item $B(next^{a_k\cdots a_i}(x_1),\Delta(a_k\cdots a_i)+\varepsilon)\subseteq  {\cal S}_1$ and 
$B(next^{b_k\cdots b_i}(x_2),\Delta(b_k\cdots b_i)+\varepsilon)\subseteq  {\cal S}_2$,\\
and

\item $B(next^{a_k\cdots a_1}(x_1),\Delta(a_k\cdots a_1)+\varepsilon)\subseteq  R_1$ and 
$B(next^{b_k\cdots b_1}(x_2),\Delta(b_k\cdots b_1)+\varepsilon)\subseteq  R_2$.
%
\end{enumerate}

\vspace*{2mm}

Then 
we have for all~$z=(z_1,z_2)\in B(x_1,\varepsilon)\times B(x_2,\varepsilon)$,
and all $i\in\{1,\dots,k\}$:\\

\begin{itemize}
\item $X_{(k-i+1)\tau,z_1}^{a_k\cdots a_i}\in  S_1$ and 
$X_{(k-i+1)\tau,z_2}^{b_k\cdots b_i}\in  S_2$ \ \ \ \hspace*{\fill} (safety), 

and

\item 
$(X^{a_k\cdots a_1}_{k\tau,z_1},X^{b_k\cdots b_1}_{k\tau,z_2})\in R$.
\hspace*{\fill}  ($k$-capture).
\end{itemize}
\end{theorem}

\ \\

%
%
%
%
%
%

\begin{example} (Tag-Chase game)
Let us consider Example~\ref{ex:2}
with $V_1=2$, $V_2=1$, $S=[-2,2]^4$,
$m_1=m_2=|A|=|B|=6$,
$A=B=\{\pm \pi/3, \pm \pi/2, \pm 2\pi/3\}$.
Let the target $R$ be defined by: $R=\{(z_1,z_2)\in \mathbb{R}^4:  \| z_1 - z_2 \| \leq 0.7 \}$.

Let 
$N=10$ the number of nodes in each dimension, 
$\varepsilon=0.31$, $\tau=0.2$. 
One can check that conditions 1 and 2 
of Theorem~\ref{th:capture} are satisfied for $k=1$.
Applying the corresponding strategy to
$z=(z_1,z_2)$ with $z_1=(1.7,1.7)$, $z_2=(1.5,1.0)$,  
it can be shown
that the controlled trajectory, after 66 steps,
reaches the state $z_1=(1.7,1.04)$, $z_2=(1.67,0.9)$ which belongs to the target $R$.
See the controller simulation Figure~\ref{fig:tag_chase} where the player 1 (pursuer) is in blue, and the player 2 
(evader)
in red.
For this initial state, one can observe that, at step 68, a limit cycle of length 16 is reached (see Figure \ref{fig:simu_cycle}). Note that, for other initial states, the length of the limit cycle may be different, and is often 2.
The experiment takes $534$ seconds of CPU time. 
%

\begin{figure}
\includegraphics[width=0.45\textwidth, clip, trim = 1cm 6cm 1cm 7cm]{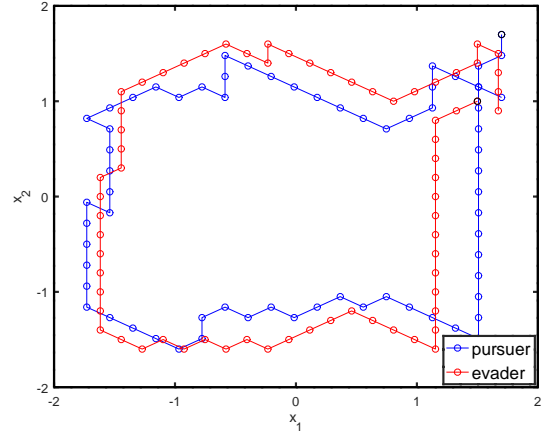}
\caption{Simulations of the tag-chase game. The initial states are $z_1=(1.7,1.7)$, $z_2=(1.5,1.0)$. The capture states
are  $z_1=(1.7,1.04)$, $z_2=(1.67,0.9)$.}
\label{fig:tag_chase}
\end{figure}

\begin{figure}
\includegraphics[width=0.43\textwidth, clip, trim = 1cm 6cm 1cm 7cm]{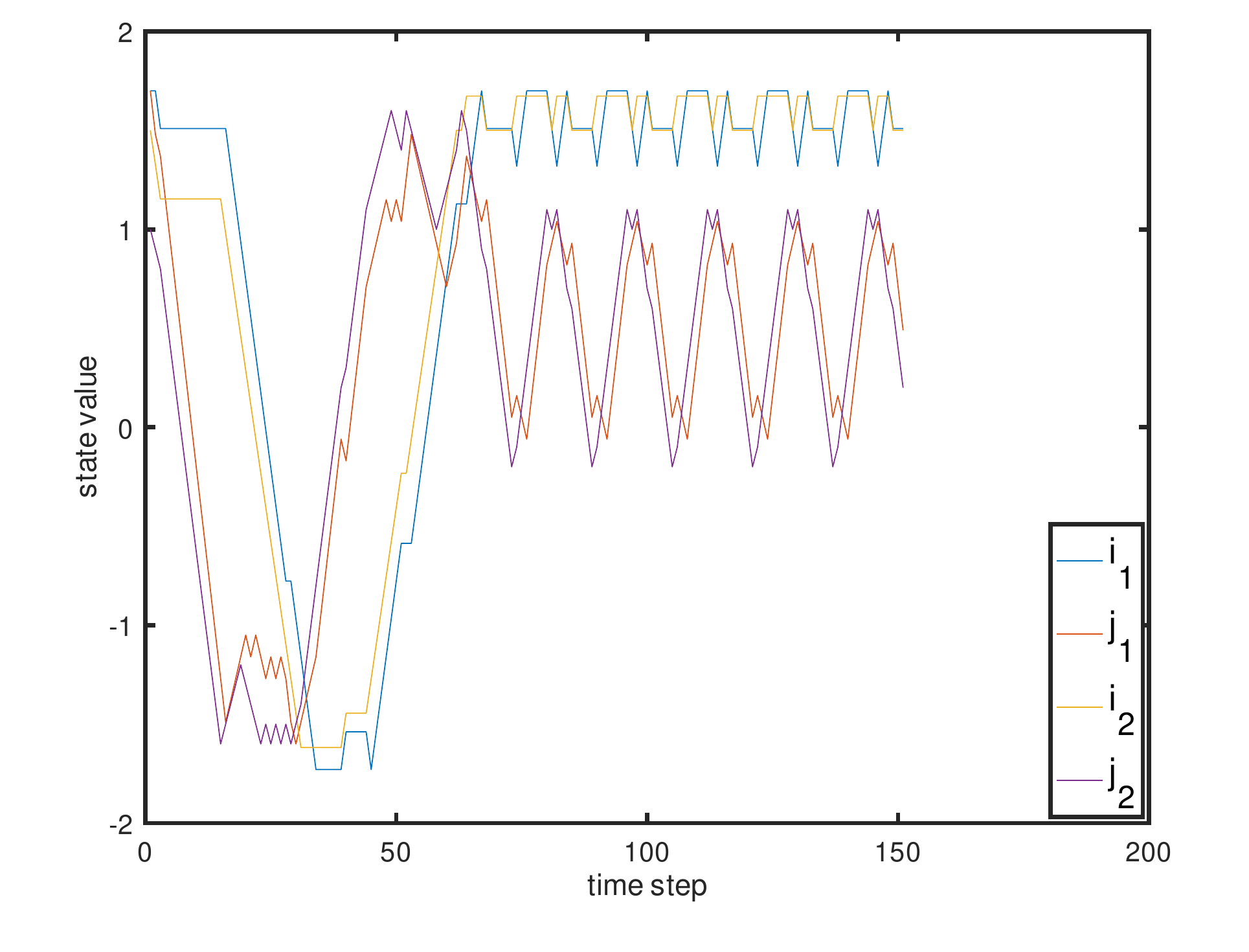} \\
\includegraphics[width=0.43\textwidth, clip, trim = 1cm 6cm 1cm 7cm]{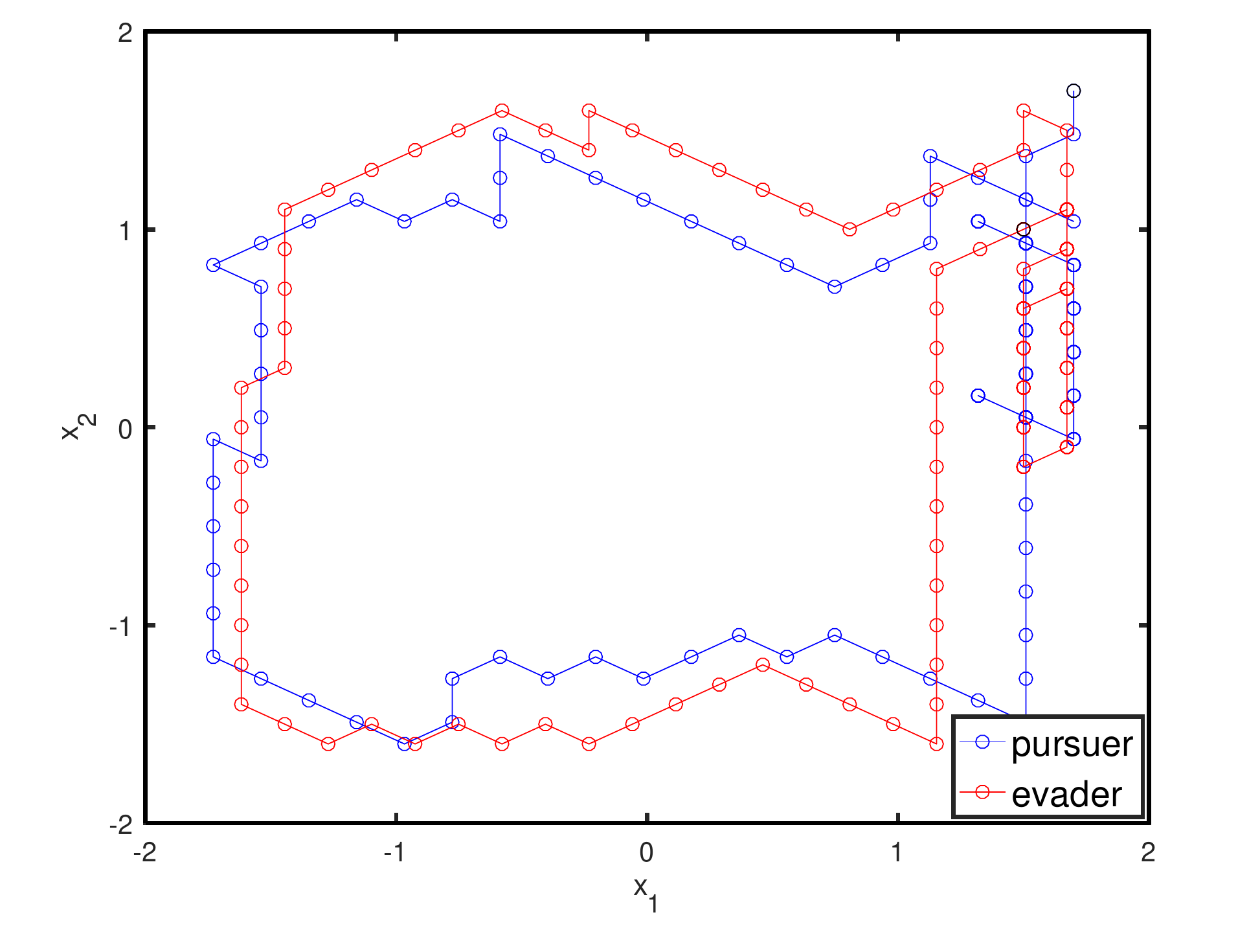}
\caption{Simulations of the tag-chase game. The initial states are $z_1=(1.7,1.7)$, $z_2=(1.5,1.0)$. At step 68, a limit cycle of length 16 is reached.}
\label{fig:simu_cycle}
\end{figure}

\end{example}

\section{Final Remarks}


We have presented a new SL-based method for synthesizing
a provably safe finite-time horizon control.
We have illustrated the interest of the method on a classical example (2-tanks) and shown how to extend it to stochastic ODEs and differential games.
The potential application of such  methods 
to Model Predictive Control has been pointed
in \cite{c4}.

A defect of our method is that, in order to satisfy the sufficient conditions of Theorem~\ref{th:MINIMATOR}, one may have to decrease the cell size $\varepsilon$ too much, thus making the number of cells explode, as often in 
SL methods.
In this case, methods using symbolic reachability analysis, such as 
in \cite{c3,c4,c5}, may be more efficient.
A comparative experimental work between the two kinds of method is planned for future work.







\bibliographystyle{plain}
\bibliography{biblio}

\newpage
\section{Appendix: error bounding for the tamed Euler scheme}

\subsection{Assumptions}
The symbol $\norm{\cdot}$ 
denotes the Euclidean norm on $\mathbb{R}^n$.
The symbol $\langle \cdot, \cdot\rangle$ denotes the scalar product of two vectors of $\mathbb{R}^n$. Given a point $x\in\mathbb{R}^n$ and a positive real
$r>0$, the ball $B(x,r)$ of centre $x$ and radius $r$ is the set
$\{y\in\mathbb{R}^n \ |\ \norm{x-y}\leq r\}$.


Let $\tau\in(0,\infty)$ be a fixed real number,
let $(\Omega,\mathcal{F},\mathbb{P})$ be a probability space with normal filtration
$(\mathcal{F}_t)_{t\in[0,\tau]}$, let $n,m\in\mathbb{N} :=\{1,2,\dots\}$
let $W=(W^{(1)},\dots,W^{(m)}): [0,R]\times \Omega \rightarrow \mathbb{R}^m$
be an $m$-dimensional standard 
$(W_t)_{t\in[0,\tau]}$-Brownian motion and let $x_0:\Omega\rightarrow \mathbb{R}^n$
be an $\mathcal{F}_0/\mathcal{B}(\mathbb{R}^n)$-measurable mapping with
$\EE [\norm{x_0}^p]<\infty$ for all $p\in[1,\infty)$.
Moreover, let $f: \mathbb{R}^n\rightarrow \mathbb{R}^n$ be a continuously differentiable function
whose derivative grows at most polynomially. Formally, let us suppose the existence of
 constants $D\in \mathbb{R}_{\geq 0}$ and $q\in \mathbb{N}$ 
such that, for all $x,y\in \mathbb{R}^n$
\eq{
\label{hyp:lipsf}
\norm{f(x)-f(y)}^2\leq D\norm{x-y}^2(1+\norm{x}^q+\norm{y}^q)
\tag{H1}
}
Let $g=(g_{i,j})_{i\in\{1,\dots,d\},j\in\{1,\dots,m\}}: \mathbb{R}^n\rightarrow\mathbb{R}^{d\times m}$ be a globally Lipschitz 
continuous function: there exists $L_g\in \mathbb{R}_{\geq 0}$ 
such that, for all $x,y\in \mathbb{R}^n$
\eq{
\label{hyp:lipsg}
\norm{g(x)-g(y)}\leq L_g\norm{x-y}
\tag{H2}
}
Finally, let us suppose that f is globally one-sided Lipschitz 
with constant
$\lambda\in\mathbb{R}$:
\eq{ 
\label{hyp:os-lips}
\exists \lambda\in\mathbb{R}\  \forall \bx,\by\in \mathbb{R}^n:\ 
\langle \bff(\by)-\bff(\bx), \by-\bx \rangle \leq \lambda\, \|\by-\bx\|^2
\tag{H3} 
}

Then consider the Stochastic Differential Equations (SDE):
\eql{
\d X_t = f(X_t)\d t+g(X_t)\d W_t,
\quad \quad X_0 = x_0
\label{eq:sde2}
}

for~$t\in[0,\tau]$.
The drift coefficient $f$ is the infinitesimal mean of the process
$X$ and the diffusion coefficient $g$ is the infinitesimal standard deviation
of the process $X$. Under the above assumptions, the SDE~\eqref{eq:sde2}
is known to have a unique strong solution. More formally, there exists an
adapted stochastic process $X:[0,\tau]\times \Omega\rightarrow \mathbb{R}^n$
with continuous sample paths fulfilling
\eq{
X_{t,x_0}=x_0+\int_0^t f(X_s) \d s +\int_0^t g(X_s) \d W_s
}
for all $t\in [0,\tau]$ $\mathbb{P}$-a.s.
(see, e.g., \cite{O02}).

We denote by $X_{t,x_0}$ the solution of Equation~\eqref{eq:sde2} at time~$t$
from initial condition $X_{0,x_0}=x_0$ $\PP{}$-a.s., in which $x_0$ is a random 
variable that is measurable in $\mathcal{F}_0$.

\begin{remark}\label{remark:1}
Constants $\lambda$, $L_g$ and $D$ can
be computed using (constrained) optimization algorithms (see \cite{c13}).
\end{remark}

\subsection{Tamed Euler scheme}

The standard time-discrete {\em tamed Euler scheme} is defined as a follows.
Let $\underline{X}_{n,z}^N:\Omega\rightarrow \mathbb{R}^d$, 
\begin{equation}
\begin{aligned}
\underline{X}_{n+1,z}^N=  \underline{X}_{n,z}^N+\frac{\frac{\tau}{N}\cdot f(\underline{X}_{n,z}^N)}{1+\frac{\tau}{N}\cdot\norm{f(\underline{X}_{n,z}^N)}} \\
 + g(\underline{X}_{n,z}^N)(W_{\frac{(n+1)\tau}{N}}-W_{\frac{n\tau}{N}})
\end{aligned}
\label{HJK(8)}
\end{equation}
for all $n\in\{0,1,\dots,N-1\}$ and all $N\in\mathbb{N}$.  In this method
the drift term $\frac{\tau}{n}\cdot f(\underline{X}_{n,z}^N)$ is ''tamed'' by the factor
$1/(1+\frac{\tau}{N}\cdot \norm{f(\underline{X}_{n,z}^N)})$ for $n\in\{0,1,\dots,N-1\}$ and
$N\in\mathbb{N}$ in~\eqref{HJK(8)}. 

A time continuous interpolation of the tamed Euler scheme (introduced in \cite{HJK11}) is written as follows.
Let $\tilde{X}^N_{z}:[0,\tau]\times\Omega\rightarrow\mathbb{R}^n$, $N\in\mathbb{N}$,
be a sequence of stochastic processes given by
\eql{\label{eq:plin-interp}
\tilde{X}_{t,z}^N=\tilde{X}_{n,z}^N+\frac{(t-n\tau/N)\cdot f(\tilde{X}_{n,z}^N)}{1+\tau/N\cdot \norm{f(\tilde{X}_{n,z}^N)}}
+ g(\tilde{X}_{n,z}^N)(W_t-W_{\frac{n\tau}{N}})
}

for all $t\in[\frac{n\tau}{N},\frac{(n+1)\tau}{N}]$, $n\in\{0,1\dots,N-1\}$ and all $N\in\mathbb{N}$.
Note that $\tilde{X}^N_{t,z}:[0,\tau]\times \Omega \rightarrow \mathbb{R}^n$ is
an adapted stochastic process with continuous sample paths for every
$N\in\mathbb{N}$.

\begin{lemma}
\label{lemma:4.3} 
Let us suppose (H1) (H2) and (H3).
Let the setting in this section be fulfilled,
and $z:\Omega\rightarrow \mathbb{R}^n$ be
an $\mathcal{F}_0/\mathcal{B}(\mathbb{R}^n)$-measurable mapping with
$\EE [\norm{z}^p]<\infty$ for all $p\in[1,\infty)$.
Then, for any even integer~$r\geq 2$, there exist
two constants $E_{r,z}$ and~$F_{r,z}$
such that
$$\sup_{0\leq t\leq \tau} \EE \norm{\underline{X}_{t,z}-\tilde{X}_{t,z}}^r\leq (\Delta_t)^{\frac{r}{2}} (E_{r,z}(\Delta_t)^{\frac{r}{2}}+F_{r,z}d).$$
with $\Delta_t=\tau/N$ and:

%
$E_{r,z}= 2^r(\norm{f(0)}^r+D2^{\frac{r+1}{2}}$\\
%
\hspace*{\fill}$ (1+\EE \sup_{0\leq t\leq \tau}\norm{\underline{X}_{t,z}}^{qr})^{\frac{1}{2}}
(\EE\sup_{0\leq t\leq \tau}\norm{\underline{X}_{t,z}}^{2r})^{\frac{1}{2}})$,\\

$F_{r,z}=2^r(\norm{g(0)}^{2r}+L_g^r\EE\sup_{0\leq t\leq \tau}\norm{\underline{X}_{t,z}}^{\frac{r}{2}}).$
\end{lemma}

\begin{remark}
Constants $E_{r,z}$ and $F_{r,z}$ are computed using the  constants $\lambda$ and $L_g$
(see Remark~\ref{remark:1}), and the expected values of $\underline{X}_{t,z}$
at each time $t=0,\Delta t, 2\Delta t, \dots, N\Delta t$.
These expected values are computed using a Monte Carlo method (by averaging here
the value of~$10^4$ samplings).
\end{remark}

\subsection{Mean square error bounding}
The following Theorem holds for SDE \eqref{eq:sde2}. This corresponds to a stochastic version of Theorem~1 of \cite{c13}, showing that a similar result holds on average, using the tamed Euler method of \cite{HJK11}.
It is an adaptation of Theorem 4.4 in \cite{HMS02}.
\begin{theorem}\label{th:stochastic}
Given the SDE system \eqref{eq:sde2} satisfying (H1)-\eqref{hyp:lipsg}-\eqref{hyp:os-lips}. Let~$\delta_0\in\mathbb{R}_{\geq 0}$.
Suppose that $z$ is a
random variable on $\mathbb{R}^n$ such that
$$\EE[\norm{x_0-z}^2]\leq \delta_0^2.$$
Then, 
we have, for all $\tau\geq 0$:
%
\eq{
\EE[ \sup_{0\leq t\leq \tau}\norm{X_{t,x_0} - \tilde{X}_{t,z}}^2 ] \leq \delta_{\tau,\delta_0}^2,
}
with $\delta_{\tau,\delta_0}^2 := \beta(\tau)e^{\gamma \tau}$, where:\\
%

$\gamma=2(\sqrt{\Delta_t}+2\lambda+L_g^2+128 L_g^4)$, and
\begin{equation}
\begin{aligned}
& \beta(\tau) = 2\delta_0^2
+2\tau\Delta_tL_g^2(1+128L_g^2)(F_{2,z}d+E_{2,z}\Delta_t)\\
& +4\tau\sqrt{\Delta_t}D(F_{4,z}d+E_{4,z}\Delta_t^2)^{\frac{1}{2}}\\
& (1+4 \EE\sup_{0\leq t\leq \tau}\norm{\underline{X}_{t,z}}^{2q}+ 4\EE \sup_{0\leq t\leq \tau}\norm{\tilde{X}_{t,z}}^{2q})^{\frac{1}{2}}.
\end{aligned}
\label{eq:th2}
\end{equation}

with 
$\Delta_t= \tau/N$.

\end{theorem}

\end{document}